\documentclass[a4paper,10pt]{article}

\usepackage[numbers,sort&compress]{natbib}

\usepackage[english]{babel}

\usepackage[utf8]{inputenc}
\usepackage[T1]{fontenc}

\usepackage{mathpazo}
\usepackage{amsmath,amssymb}
\usepackage{bm} 
\usepackage{bbm} 
 
\usepackage{textcomp}

\usepackage[final]{graphicx}

\newcommand{\ket}[1]{\vert{#1}\rangle} 
\newcommand{\bra}[1]{\langle{#1}\vert} 
\newcommand{\bracket}[2]{\langle{#1}\vert{#2}\rangle} 
\newcommand{\proj}[1]{\ket{#1}\!\bra{#1}}
\newcommand{\mean}[1]{\langle #1 \rangle}
\DeclareMathOperator{\Tr}{Tr}

\newcommand{\abs}[1]{\left|#1\right|} 

\newcommand{\beq}{\begin{equation}}
\newcommand{\eeq}{\end{equation}}
\newcommand{\up}{\uparrow}
\newcommand{\down}{\downarrow}
\newcommand{\one}{\mathbbm{1}}

\renewcommand{\vec}[1]{\boldsymbol{#1}}

\begin{document}

\title{Decoherence in the chemical compass\\ \Large The role of decoherence for avian magnetoreception}

\author{Markus Tiersch%
\thanks{Electronic address: markus.tiersch@uibk.ac.at}~
\thanks{Institute for Quantum Optics and Quantum Information, Austrian Academy of Sciences, Technikerstr.~21A, A--6020 Innsbruck, Austria;
Institute for Theoretical Physics, University of Innsbruck, Technikerstr.~25, A--6020 Innsbruck, Austria} \and 
 \and Hans J. Briegel\footnotemark[2]
}
\date{}


\maketitle

\begin{abstract}
Contrary to the usual picture that decoherence destroys quantum properties and causes the quantum-to-classical transition, we argue that decoherence can also play a constructive role in driving quantum dynamics and amplifying its results to macroscopic scales.
We support this perspective by presenting an example system from spin chemistry, which is also of importance for biological systems, e.g.\ in avian magnetoreception.
\end{abstract}

\section{Introduction}

In his famous 1935 paper~\cite{Schroedinger}, Erwin Schr\"{o}dinger described a notorious ge\-dan\-ken experiment with a cat whose state of being alive or dead becomes entangled with the state of a single radioactive atom that triggers the release of a poisoning substance upon its decay. The main purpose of his argument was to illustrate the apparently absurd implications of the quantum mechanical superposition principle when it is applied to the macroscopic world of everyday objects. There have been many discussion of this gedanken experiment (for a review see e.g. Zurek~\cite{Zurek}), most of which are trying to show, one way or the other, why coherent superpositions of macroscopic objects of the type described by Schr\"{o}dinger cannot be observed in real life. The mechanism that is made responsible for this impossibility is accounted for by quantum mechanics itself, and is usually termed (environmentally induced) decoherence. 
Decoherence arises due to interactions of the quantum system of interest with environmental degrees of freedom which are not controlled by the experimenter. With increasing size of the quantum system, measured e.g.\ by its numbers of degrees of freedom, its state usually becomes more and more fragile and susceptible to disturbances. The time after which quantum superpositions of larger systems turn into incoherent probabilistic mixtures, thereby losing their capability to interfere, becomes shorter and shorter with increasing system size, and vanishes in the macroscopic limit. This loss of the interference terms, i.e.\ of the quantum coherence of the quantum system, is sometimes termed quantum-to-classical transition and it suggests the existence of a certain (not very sharply defined, but otherwise persistent) borderline between the quantum and the classical world.%
\footnote{The appearance of a classical world is ultimately caused by our inability to keep track of the correlations between the system's and the environment's degrees of freedom.}

Decoherence, in this picture, plays the role of a destructive mechanism that ``washes out'' quantum effects when the system size increases. It explains why we usually do not observe interference of distinct quantum states, i.e., the absence of quantum coherence, on a macroscopic level, and why ``Schr\"{o}dinger's cat'' will be either alive or dead, but not both at the same time. 
It is a different question to what extent the (classical) description of things and processes that go on at the macroscopic level are still dependent on a proper (quantum) description of processes on the quantum level. Surely, the decay of a single radioactive particle may have a severe effect for macroscopic objects, including the health of human beings. But, to understand this effect, the radioactive decay can effectively be described by a classical random event. In this sense (and for this specific example) there is a conceptually closed phenomenology for the classical world of the cat. It is however conceivable that the behavior of macroscopic things and beings, such as a cat, may depend in a much more direct way on the explicit quantum nature of processes on the molecular level.   
One example where this might indeed be the case lies in the context of bird navigation. Certain birds, including the European robin, have the remarkable ability to orient themselves, during migration, with the help of the Earth's magnetic field~\cite{Wiltschko1,Wiltschko2,Mouritzen01,PhysToday}. Responsible for this ``magnetic sense'' of the robin, according to one of the main hypotheses, seems to be a molecular process called radical pair mechanism~\cite{Schulten2,SchultenRev} (also, see \cite{Ritz00,Rodgers09} for reviews that include the historical development and the detailed facts leading to the hypothesis). It involves a photo-induced spatial separation of two electrons, whose spins interact with the Earth's magnetic field until they recombine and give rise to chemical products depending on their spin state upon recombination, and thereby to a different neural signal. The spin, as a genuine quantum mechanical degree of freedom, thereby controls in a non-trivial way a chemical reaction that gives rise to a macroscopic signal on the retina of the robin, which in turn influences the behavior of the bird. When inspected from the viewpoint of decoherence, it is an intriguing interplay of the coherence (and entanglement) of the initial electron state and the environmentally-induced decoherence in the radical pair mechanism that plays an essential role for the working of the magnetic compass.

In this paper we will analyze the radical pair mechanism from the perspective of decoherence theory and the dynamics of open quantum systems. The radical pair model is the underlying mechanism for the chemical compass, which in turn is believed to be the molecular basis for avian magnetoreception.
The present work is not meant to be an in-depth analysis of the avian compass itself. (For recent reviews, see \cite{Ritz00,Rodgers09} for example.)
Rather than studying the detailed molecular realization of the avian compass, we will restrict our analysis to generic models of the radical pair mechanism that capture the essential features.
Thereby we identify to what extent fundamental quantum-mechanical processes, decoherence in particular, can affect the behavior of macroscopic and possibly living things.

\section{Decoherence in subsystems}

When describing the states of physical systems, it seems that on the microscopic scale the quantum formalism should be used, whereas at larger scales in most cases classical physics is sufficient to capture the full phenomenology. At somewhat intermediate scales the quantum properties of the considered physical system ``fade away'' when a certain size is reached, and thus one changes the formalism that is used to describe the physics. This regime is usually identified with the term \emph{quantum-to-classical transition}.

The loss of quantum properties manifests itself in the vanishing quantum coherence in the system, which can be defined as the inability to exhibit superpositions and interference phenomena. Technically, quantum coherences are contained in the off-diagonal matrix elements of the density matrix of a quantum system.

In order to investigate how quantum systems turn into systems whose phenomenology is contained within a classical model, it is necessary to discuss how quantum coherences disappear for large quantum systems. This process is termed \emph{decoherence}.

Decoherence naturally appears within the formalism of quantum physics when different \emph{subsystems} are considered. The necessity of introducing the concept of a scope arises from a simple fact: A realistic treatment of quantum systems cannot ignore that every quantum object is in contact with its surrounding, however small their interaction may be. The quantum object of interest can therefore not be treated as fully independent from its surroundings. Consequently, and in order to arrive at meaningful statements about the system of interest, it is necessary to introduce a scope that only extends over the system, in fact, defining the system.
Laboratory experiments that demonstrate quantum effects usually first need to isolate the system from its environment in order for the desired quantum effects to prevail on long enough time scales, as to be visible and demonstrated. Although experimental efforts result in ever better isolation, it will only be possible to converge towards perfection. To make things worse, such procedures typically become increasingly harder for larger and larger quantum systems.

More formally, it is first of all necessary to identify and distinguish a system from its environment, in order to specify the object of concern whose quantum (or classical) properties are of interest. This could be the electronic state of an atom trapped in a vacuum chamber, the spin-property of an electron, or the register of a quantum computer implementation. The rest of the universe then constitutes the system's environment, in principal. However, it is usually sufficient to include into the environment those degrees of freedom of the environment that couple most strongly to the system, because they are the ones that cause the arising effects during short timescales. For the quantum systems given above, these environmental degrees of freedom could be the surrounding electromagnetic field, either in the vacuum or in a thermal state, stray magnetic fields, or the control apparatus of the quantum computer.

Once system and environment have been identified, their interaction dynamics on the quantum level results in a decoherence process for the system degrees of freedom. The decoherence thus arises naturally for every quantum system on the quantum level, once the interaction with an environment is taken into account, but the focus is restricted to the system alone. In this step, restricting the focus to the system alone, thereby ignoring the environment, leads to a loss of information also for the system degrees of freedom because of their interaction. The procedure of ignoring some part of a composite quantum will therefore need some further elaboration. It is this limited scope that ultimately gives rise to a loss of coherences as they become ``redistributed'' in the composite system-environment quantum system. Since the environment is virtually of unlimited size (strictly, the rest of the universe), a ``rephasing'' and revival of coherences back in the system is prevented on finite timescales.

\bigskip
In the following we review some of the textbook knowledge in order to detail the statements from above using quantum physics notation. We follow the established formalism how decoherence arises in so-called open quantum system~\cite{Breuer,Gardiner}. The task of identifying the degrees of freedom of interest, the \emph{system}, and its environment mathematically amounts to defining the respective Hilbert spaces. That is, one defines a partitioning of the entire composite Hilbert space (comprising the universe) into the system and the environment (rest of the universe), by establishing a tensor product structure:
\beq
\mathcal{H} = \mathcal{H}_S \otimes \mathcal{H}_E.
\eeq
Here, $\mathcal{H}_S$ and $ \mathcal{H}_E$ denote the Hilbert spaces of system and environment, respectively.

In order to witness the decoherence process in detail, it is necessary to first specify an initial state and then to observe the subsequent dynamics. For demonstration purposes, here we assume that system and environment each start in a pure quantum state. That is, the initial state of the entire system is given by the pure product state
\beq \label{eq:initialstate}
\ket{\Psi(t_0)} = \ket{\psi_S(t_0)} \otimes \ket{\psi_E(t_0)}.
\eeq
This assumption may easily be relaxed to more general initial state such as products of mixes states, \mbox{$\rho(t_0)=\rho_S(t_0)\otimes\rho_E(t_0)$}, and even initially correlated states, i.e.\ non-product states of system and environment, can be treated by including them as the result of a dynamic process starting from an initial product state.
We leave aside the discussion about which quantum state should be assumed for the entire universe, as for practical purposes the system's immediate environmental degrees of freedom are of importance. Weakly and indirectly coupled, or more distant degrees of freedom will not influence the dynamics on sufficiently short timescales.

Focusing on the dynamics that cause the decoherence process, we now turn to the system-environment interaction. Once the partition into system and environment has been defined, the complete Hamiltonian can be grouped into a part that only affects the system, $H_S$, one that only acts on the environment, $H_E$, and a part that contains operators, which act on both systems, i.e.\ the interaction Hamiltonian $H_{int}$. We thus write
\beq
H = H_S \otimes \one_E + \one_S \otimes H_E + H_{int},
\eeq
where the trivial action on system and environment has been included in form of the identity operator for utmost clarity.
The time evolution of the composite system is given by the formal solution to the Schr\"{o}dinger equation:
\beq
\ket{\Psi(t)} = U(t) \ket{\Psi(t_0)} = e^{-i H t/\hbar} \ket{\Psi(t_0)}.
\eeq
While the initial state of system and environment is a product state, cf.~\eqref{eq:initialstate}, this is no longer the case at later times in general. Instead the state will assume the form
\beq
\ket{\Psi(t)} = \sum_{i,j} c_{ij} (t) \ket{\phi_i} \otimes \ket{e_j},
\eeq
where $\{\ket{\phi_i}\}$ and $\{\ket{e_j}\}$ are sets of basis vectors in the Hilbert spaces of system and environment, respectively. The expansion coefficients $c_{ij}$ are non-zero, in general.

At this point, we have only extended the quantum description of our system of interest to even more degrees of freedom and thereby included the environment. In order to arrive at the effective dynamics for the system only, it is necessary to ``get rid'' of the environmental degrees of freedom, however, without dispensing with their influence on the system state.
The appropriate operation consists in performing the \emph{partial trace} over the environmental degrees of freedom.

The act of limiting the scope to the system degrees of freedom by ``tracing out'' the environmental ones is the unique procedure to arrive at states and statistics of observables for a part of a composite systems. For a measurement that is only carried out on a subsystem, one needs to sum over all measurement results in the composite system that yield the same measurement outcome for this subsystem. Statistically this amounts to calculating a marginal probability distribution.
For example, a measurement on the system that asks if the system is in the state $\ket{\phi}$ returns, when lifted to system and environment, a positive result for any state of the form $\ket{\Phi}=\ket{\phi}\ket{e_j}$ where $\ket{e_j}$ is any state of the environment.

More generally, the statistics for measurements of any observable on the system is fully contained in a system's quantum state $\rho_S$. In the case of pure quantum states it is the projector $\rho_S=\proj{\psi_S}$, in fact this is the only case for which $\Tr \rho_S^2=1$ holds.
Given a mixed state $\rho$ of system an environment, the statistics of an observable $A_S$ on the system, e.g.\ the expectation value, is given by the statistics of the trivial embedding of the operator to the composite system, i.e.\ by $A_S\otimes\one_E$. The expectation value of $A_S$ for the system is therefore extended to the calculation of an expectation value in the composite space, and calculated as follows:
\begin{align}
\mean{A_S} &= \Tr {A_S\otimes\one_E \rho} = \sum_{i,j} \bra{\phi_i}\bra{e_j}A_S\otimes\one_E \rho \ket{\phi_i}\ket{e_j} \nonumber \\
&= \sum_i \bra{\phi_i} A_S \rho_S \ket{\phi_i} = \Tr_S A_S \rho_S.
\end{align}
Here, the trace over system and environment is taken apart and regrouped to yield the expectation value for the system with respect to its density operator
\beq
\rho_S = \sum_i \bra{e_i}\rho\ket{e_i} = \Tr_E \rho.
\eeq

Adopting this perspective allows us to identify that the decoherence of a quantum system, i.e.\ the dynamics that take an initially pure (superposition) state of the system with coherences ($\proj{\psi_S}$) to a mixed state with decayed coherences and a decreased purity ($\rho_S$ with $\Tr \rho_S^2 <1$), i.e.\ a contribution of ``classical ignorance'' about the state of the system.

Let us illustrate this with an example of an atom (system) and a detector (environment). Initially the atom is assumed to be excited and the detector is in a neutral/null position. After some time of interaction, the composite system of atom and detector would be described as
\beq
\ket{\Psi} = \alpha \ket{\textsf{excited}}\ket{\textsf{neutral}} + \beta \ket{\textsf{decayed}}\ket{\textsf{affected}}.
\eeq
Tracing over the detector degrees of freedom, the atomic state is mixed in general and reads in matrix notation (using basis $\ket{\textsf{excited}}$, $\ket{\textsf{decayed}}$ of the atom)
\beq
\rho_\textsf{atom} =
\begin{pmatrix} 
\abs{\alpha}^2 & c \\
c^* & \abs{\beta}^2
\end{pmatrix},
\eeq
where the coherences still exhibited by the atom are determined by how much the detector has been affected,
\begin{align}
c &= \alpha\beta^* \Tr_\textsf{detector} \ket{\textsf{neutral}}\bra{\textsf{affected}} \nonumber \\
&= \alpha\beta^* \bracket{\textsf{affected}}{\textsf{neutral}},
\end{align}
i.e.\ they depend on how different the ``environment states'' are, here that of the detector.
If the detector would have only been weakly coupled to the atom, and thus only affected by the decay minutely, e.g.\ the detector gained some small momentum due to the atomic decay products, but not enough to click, the detector states maintain a large overlap $\bracket{\textsf{affected}}{\textsf{neutral}}\approx 1$.
However, if the detector really clicked, i.e.\ $\ket{\textsf{affected}}=\ket{\textsf{clicked}}$, the overlap is zero by definition $\bracket{\textsf{clicked}}{\textsf{neutral}}=0$.

\section{Radical Pair Mechanism}

After having collected the essential notions and how decoherence naturally emerges from an open quantum system approach, we proceed to identify how decoherence can play a constructive role in a concrete scenario: the radical pair mechanism, which occurs in molecular systems of spin chemistry and in biomolecules.


The theory of the radical pair mechanism was developed during the 1970s (see \cite{Steiner} for an historical account of the development) to explain how light-induced chemical reactions can exhibit a dependence on an external magnetic field. Since then these chemical reaction that involve electron spins have been extensively studied by spin chemists in molecular systems~\cite{Steiner,Salikhov,Nagakura}. It is the light- and magnetic-field-dependence that makes this chemical reaction mechanism one of the main hypotheses for the ability of animals such as migrating birds to sense the magnetic field~\cite{Schulten2,SchultenRev,Ritz00,Rodgers09,Wiltschko02,Ritz04,Mouritzen04,Maeda,photolyase,Ritz09,Solovyov,Mouritzen11}.

\begin{figure*}
    \centering
	\includegraphics{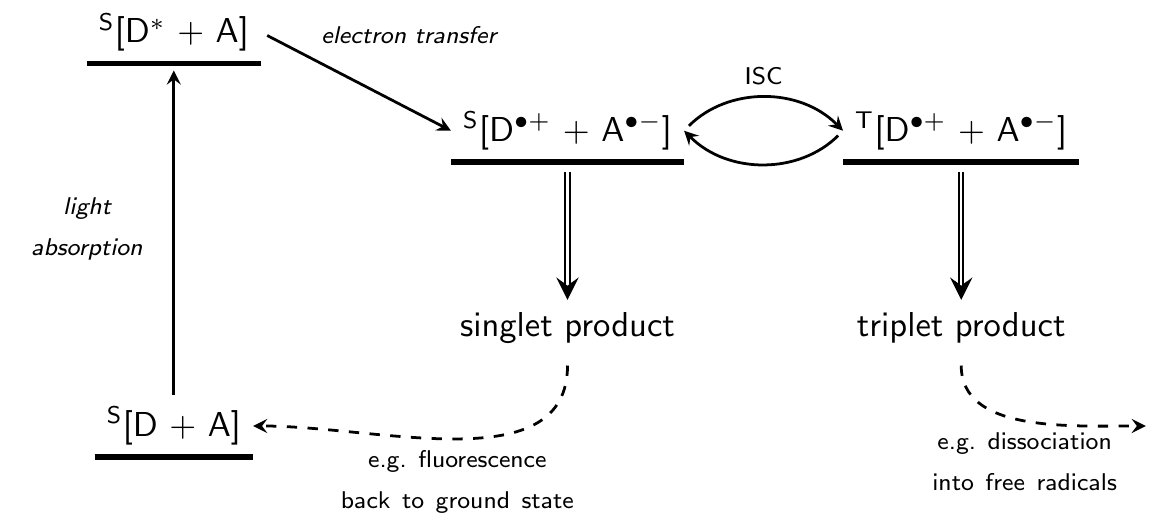}
    \caption{Scheme of the radical pair mechanism. Donor (D) and acceptor (A) molecules form a correlated spin radical pair after a light-induced electron transfer. Due to magnetic fields the spin character of the radical spins is interconverted between singlet and triplet states, the so-called inter-system crossing (ISC).}
    \label{fig:RPM}
\end{figure*}


A general scheme depicting the radical pair mechanism is shown in fig.~\ref{fig:RPM}. For concreteness, we will focus on the variant of radical pair molecules in solution, and we shall provide a qualitatively simplified discussion of the technical details of the chemistry. For the latter, the reader is asked to refer to the relevant chemistry literature. Usually there are two molecule species involved that play the role of electron donor and acceptor. After light absorption, the electronic state of one of the molecules will be in an excited state, e.g.\ one of the electrons in the donor molecule is promoted from the highest occupied molecular orbital (HOMO) to the lowest unoccupied molecular orbital (LUMO). Subsequently, if both molecules are close enough, the electron is transferred from the donor to the acceptor molecule. Both molecules have an unpaired electron and thus form the pair of radicals, hence the name. The electron that has moved from donor to acceptor left its partner, with whom it was bound in a singlet state in the electron orbital back at the donor molecule.

After some time, the electron may be transferred back to the donor to recombine provided that both molecules are close enough. A recombination back to the ground state, however, may only occur if both electrons are in a singlet state. This imposes the spin dependence in the back-reaction, i.e.\ in the ``products'' of the chemical reaction. If both electrons do not happen to be in a singlet state, they may either recombine into a long-lived triplet excited state or not recombine at all. The two molecules are therefore either unavailable for another cycle of the reaction or diffuse apart to form free radicals. Both affect the amount singlet ``reaction product''.
Often, the recombination to the ground state involves an intermediate step, the formation of an exciplex, where both molecules form a compound structure that is bound by the electron being partially back-transferred but still localized on both molecules. The recombination from this structure back to the ground state is accompanied by the fluorescence, which is easily detected in experiments and whose intensity is proportional to the abundance of singlet products.

While the radical pair is formed, the two electron spins are on different molecules and thus spatially separated. Their separation distance is even further increased if the two radicals diffuse in solution until they eventually (probabilistically) encounter again and thereby provide a possibility for the back-reaction. Electron spin-spin interactions such as the exchange and dipolar interaction are dominant while the electrons are on the same molecule, but quickly fall-off with their separation (exponential for the exchange interaction, cubic for the dipole interaction). The dominant interaction that concerns the electron spins are therefore 1) an external magnetic field and 2) the local magnetic fields due to the hyperfine coupling to the nuclear spins of the atoms that constitute the respective molecules. Exchange and dipolar interaction are negligible in comparison, or cancel~\cite{Efimova}.
These interactions cause a dynamics of the electron spin state and are able to convert electron singlet and triplet states into each other. This process is therefore often termed inter-system crossing.

In the radical pair mechanism separated electron spins evolve dynamically due to external and internal magnetic fields. This is where an external magnetic field enters in the system dynamics. Together with a spin-dependent electron recombination that constitutes the chemical reaction step, both aspects render the chemical process magnetic field dependent.

\begin{figure}
	\centering
	\includegraphics{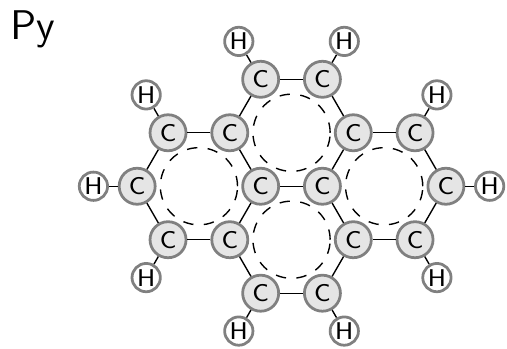}
	\hspace{1cm}
	\includegraphics{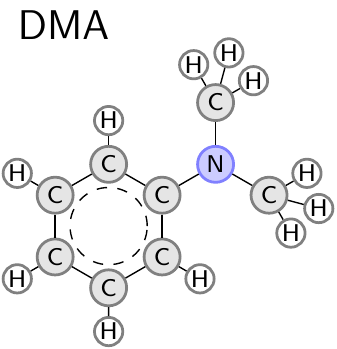}
	\caption{Schematics of molecular structures of the pyrene molecule Py (left) and the dimethylaniline DMA molecule (right) depicting all atoms: carbon (C), hydrogen (H), and nitrogen (N). Py contains 4 aromatic rings, DMA contains one aromatic ring and two methyl groups, which are connected by the nitrogen.}
	\label{fig:PyDMA}
\end{figure}

\bigskip
A well-studied example from spin chemistry is a solution of pyrene (Py) and di\-methyl\-aniline (DMA) molecules, the structure of which is given in Fig.~\ref{fig:PyDMA}. DMA serves as the donor and Py as the acceptor. Pyrene has 10 hydrogen nuclei $^1$H of spin~1/2. The DMA molecule has 11 spin-1/2 hydrogen nuclei and a nitrogen $^{14}$N nucleus of spin~1. All carbon nuclei are usually of the spin zero $^{12}$C isotope. Magnetic field effects, that is a magnetic-field dependent modulation of the fluorescence signal, have been well-studied for this pair of molecules, including isotope effects (see~\cite{Rodgers2007,Steiner} for example).

For the chemical compass hypothesis of birds, however, molecules in solution are not an option. 
One reason is that in avian magnetoreception the \emph{direction} of the magnetic field is detected, not necessarily the strength, the latter varying to a much lesser extent. Molecules in solution tumble and thus average of the field direction and therefore can only exhibit magnetic field effects when varying the field strength. Furthermore, in order to exhibit a directional dependence on the external magnetic field, the magnetic interactions between the spins on the molecule need to exhibit some direction. Otherwise there is no notion of an angle between the external magnetic field and the spin degrees of freedom of the molecules.
The radical pairs should therefore feature a sufficiently strong anisotropy of the hyperfine couplings, and moreover an ensemble of molecules should be aligned with respect to the magnetic field in order to exhibit an macroscopic \emph{angular} dependence on the magnetic field that is not averaged out with molecule orientation~\cite{Ritz00,Rodgers09,Ritz09,Solovyov}. It has been shown in~\cite{Maeda,photolyase} that it is, in principle, possible to obtain magnetic field effects for magnetic fields as weak as the Earth's magnetic field of about 50\,\textmu{}T, whereas typical hyperfine couplings and fields in lab-experiments are usually in the mT-domain (e.g.\ see~\cite{Rodgers2007}). There, the radical pair is established by two electrons on two different parts of a single larger molecule, a so-called biradical. For living organisms, the current candidate is a radical pair within a protein that exhibits a light-induced electron transfer. For the photolyase protein, the biological function of which is to repair DNA, magnetic field effects of the protein have been demonstrated in vitro~\cite{photolyase}. The candidate protein for birds, cryptochrome, is very similar to photolyase (it also belongs to the group of flavo-proteins) and is therefore suspected to be responsible for magnetic field effects that could be exploited to establish a sense for the direction of magnetic fields~\cite{Ritz00,Rodgers09,Ritz04,Mouritzen04,Ritz09,Solovyov}.

A remarkable fact regarding the magnetic-field sense exhibited by birds, if it really happens by means of the radical pair mechanism, is the difference in involved energy scales. The energy differences of the involved spin states are of the order $10^{-7}$\,eV/mT for electron spins in a magnetic field, whereas thermal energy \mbox{$k_B T=26.7$\,meV} at 310\,K, and even more, relevant chemical reactions involve energies of the order of electron volts. Despite their small difference in energy, it is the different spin states that influence and steer chemical reactions that cannot be triggered by thermal noise but require activation by light, e.g.\ green light of a wavelength of 500\,nm delivers an energy of 2.48\,eV per absorbed photon. The reason lies in the fact that these processes are non-equilibrium phenomena that take place on transient time scales faster than those of thermalization.

\bigskip
In what follows, we shall identify decoherence mechanisms at two places in the radical pair reaction scheme.
First, the interconversion and mixing of the radical singlet and triplet states is the result of a decoherence process due to hyperfine coupling of the electron spins to nuclear spins in their environment.
Second, the recombination reaction is sensitive to the spin character of the radicals and thus involves an environment that mixes singlet and triplet components of a quantum state, which also amounts to a special form of decoherence.

\section{Spin interconversion}

After the radical pair has been formed by light-induced electron transport as outlined in the previous section, we are confronted with the part of the radical pair model that is sensitive to magnetic fields.
Radical pair molecules involve two molecules, a donor and an acceptor molecule, which each carry one the two involved unpaired electrons. It is the spin state of the electron that concerns us here and that influences the subsequent chemical reaction. In addition to the unpaired electron spins, each of the molecules generally has a number of nuclear spins, and there may be an external magnetic field of artificial or natural origin.
The spin evolution is generated by the spin Hamiltonian of electron and nuclear spins:
\beq \label{eq:SpinH}
H = H_\text{zee} + H_\text{hf} + H_\text{dip} + H_\text{ex} + H_\text{nuc}
\eeq
with the individual parts as follows.

The \emph{Zeeman interaction} of each of the electron spins ($m=1,2$) with an external field $\vec{B}$ relates to the energy owing to alignment of the electronic spins in the external magnetic field:
\beq
H_\text{zee} = \sum_{m=1}^2 \mu_B g_m \vec{B}\cdot\vec{S}_m
\eeq
where $\mu_B$ is the Bohr magneton, $g_m$ is the effective $g$-factor of electron $m$, and $\vec{S}_m$ are the (dimensionless) electron spin operators, i.e.\ $\vec{\sigma}/2$ with $\vec{\sigma}$ being the vector of Pauli matrices.
The electron $g$-factor is $g_m \approx g_e\approx 2$ but varies slightly from the value for free electrons $g_e$ due to the presence of the molecule.

The \emph{hyperfine interaction} term captures the interaction between the electrons and the nuclei of the respective molecule:
\beq
H_\text{hf} = \sum_{m=1}^2 \sum_{k=1}^{N_m} \mu_B g_m \vec{S}_m \cdot \vec{\lambda}_{m,k} \cdot \vec{I}_{m,k}.
\eeq
It involves the $N_m$ nuclear spin operators $\vec{I}_{m,k}$ (again without dimension) of each of the molecules coupled to the respective electron spin. The hyperfine coupling strength between an electron spin and a nuclear spin of the same molecule is contained in the hyperfine coupling tensor $\vec\lambda_{m,k}$.
For isotropic hyperfine couplings, the $\vec\lambda_{m,k}$ reduce to scalar quantities, i.e.\ the tensors are proportional to the identity matrix.
A hyperfine coupling between an electron spin an the nuclei of the other molecule can be safely neglected in comparison.

The \emph{dipole-dipole interaction} between the electron spins is small when both molecules diffuse in the solution and are separated by a large distance, and it is therefore usually neglected. In scenarios where the electrons are fairly close such as in proteins, dipolar and exchange interaction may cancel~\cite{Efimova}. As a function of the separation distance $R$ of the electrons the dipolar interaction drops off as
\begin{equation*}
H_\text{dip} = \mathcal{O}(1/R^3).
\end{equation*}

Similarly, the \emph{exchange interaction} between electrons can be ignored for a large separation of the molecules when the electronic wave functions do not overlap anymore. It decreases exponentially, i.e.
\begin{equation*}
H_\text{ex} = \mathcal{O}\left(e^{-R}\right),
\end{equation*}
for large distances.

The \emph{interaction of the nuclear spins} with the external field and among themselves within the respective molecule are also negligible because, due to the mass difference, the gyromagnetic ratio is at least three orders of magnitude smaller than those of the electrons:
\begin{equation*}
H_\text{nuc}=\mathcal{O}\left(10^{-3}H_\text{zee}\right).
\end{equation*}

\begin{figure}
	\centering
	\includegraphics{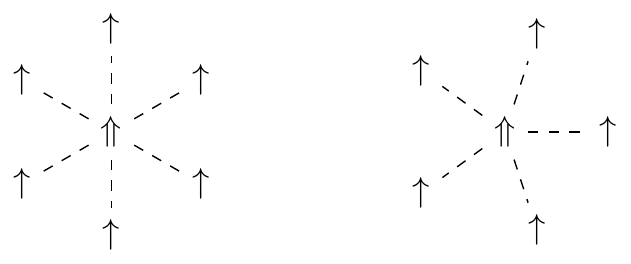}
	\caption{Graph of the topology of hyperfine interactions between the spins of a radical pair. In this example the electron spins ($\Uparrow$) are coupled to six and five nuclear spins ($\uparrow$) of the respective radical molecule, respectively. The number of nuclear spins may vary, whereas the star-graph structure is generic.}
	\label{fig:HFtopology}
\end{figure}

Focusing on the relevant, i.e.\ dominating magnetic interactions, it becomes apparent that the electron spins are not coupled to each other anymore, but subject to a local environment of nuclear spins of the respective molecule and the external magnetic field. The graph of the dominant spin interactions therefore takes the form of two stars with the electron spins in the center as depicted in fig.~\ref{fig:HFtopology}.

For the \emph{initial states}, it is usually assumed that the radical pair is formed by a fast electron transfer from a pair of covalently bond electrons, which are in a singlet state (cf.\ fig.~\ref{fig:RPM}). The usual rationale behind the spin conservation during the fast electron transfer is that magnetic interactions as in~\eqref{eq:SpinH} do not have sufficient time to change the spin state.
Although processes are known that form radical pairs differently, we adopt the standard assumption (that is also generally made for the Py-DMA radical pair, for example) and assume a singlet state $\ket{S}=(\ket{\up\down}-\ket{\down\up})/\sqrt{2}$ for the electron spins.
It should however be noted that different initial states would affect the occurring magnetic field effects~\cite{Cai2010}, and deviations from a pure initial singlet state have been observed in experiments~\cite{Maeda11}.
While we take a pure initial state for the electronic spins, the situation differs for the nuclear spins. Nuclear spin states are separated in energy by much less than what could be supplied by thermal fluctuations at room temperatures and magnetic fields below 1\,T. Under these circumstances the nuclear spin state is a thermal state, which is effectively completely unpolarized $\rho_\text{nuc}\propto\one$ at room temperature.

During the \emph{time evolution}, the electrons generally do not maintain their singlet state because each of them is coupled via all spin components ($S^x_m$, $S^y_m$, $S^z_m$) to nuclear spins and thus $[H,\vec{S}_m] \ne 0$, i.e.\ the singlet state, which is an eigenstate of the total electron spin $\vec{S}_1+\vec{S}_2$, is not an eigenstate of the Hamiltonian. Consequently, the electron-nuclear spin dynamics leads to a decrease of \emph{singlet fidelity} of the electronic spin state,
\beq
f_s(t) = \bra{S}\rho_\text{el}(t) \ket{S} = \Tr \big[ \proj{S}\otimes\one_\text{nuc} \; \rho(t) \big],
\eeq
because some triplet character is built up during the dynamics.
In accord with the previous section we identify the system with the electron spins, $\rho_S=\rho_\text{el}$, and the nuclei with the environment.

The specific topology of the hyperfine interaction, namely that both electron spins only interact with their respective private environment of nuclei, allows us to split the time evolution of the complete system into the independent time evolution of the two radicals. The only fact that links the two radicals and that requires us to treat both molecules at the same time is the initially correlated, in fact entangled, state of the electron spins.
Regarding the electronic spin state we obtain
\beq
\rho_\text{el}(t) = \Tr_\text{nuc} \left[ U(t)\; \proj{S}\otimes\frac{\one}{d_\text{nuc}} \; U^\dag(t) \right],
\eeq
where the nuclear spin environment is traced out from the time-evolved initial state of the complete system. Here, $d_\text{nuc}$ denotes the dimension of the Hilbert space of nuclear spins, i.e.\ $N_1 N_2$ times the product of the multiplicities of all nuclear spins.
Since the Hamiltonian is a sum of Hamiltonians of the magnetic interaction on the respective molecules, $H=H_1+H_2$, which commute with one another, the time evolution operator factors into those of the respective molecules. Note that the tensor product in $U(t)=U_1(t)\otimes U_2(t)$ with $U_i(t)=\exp(-iH_i t/\hbar)$ is with respect to the molecules, i.e.\ spins $\vec{S}_1$ and all $\vec{I}_{1,k}$ vs.\ $\vec{S}_2$ and all $\vec{I}_{2,k}$. The tensor product of the initial state, however, is with respect to electron and nuclear spins, i.e.\ spins $\vec{S}_{1,2}$ vs.\ all $\vec{I}_{m,k}$.
After taking the partial trace over the nuclear spins of the respective molecules, and utilizing that the initial state of the electron spin system and the nuclear spin environment is a product state, we can formally identify the following dynamical maps (completely positive trace-preserving maps)~\cite{Cai2010}:
\beq
\rho_\text{el}(t) = \mathcal{M}_1(t) \otimes \mathcal{M}_2(t) \; \proj{S}.
\eeq
Each of the local dynamical maps $\mathcal{M}_i(t)$ contains the electron spin evolution due to the nuclear spin environment of the respective molecule.

This is the place where we are confronted with decoherence dynamics. The electron spins start out in a singlet state, an entangled state, and start losing their coherence and thus entanglement due to interaction with local environments of nuclear spins. The dynamics that the spin state of the electrons undergoes is generally not unitary anymore owing to the coupling with the nuclear spins, but rather starts entangling nuclei and electrons. This leads to a decrease in purity of the electron spin state and thus to an incoherent admixture of electron spin states that are orthogonal to the singlet, namely triplet components.

If there are no nuclear spins or the hyperfine couplings vanish, the electron spins evolve simply due to an external magnetic field with $H=\mu_B g \vec{B}\cdot(\vec{S}_1+\vec{S}_2)$. (Differences in the effective $g$-factor are usually small in bio-organic systems with variations $\Delta g/g \lesssim 10^{-4} \ldots 10^{-3}$~\cite{MoebiusSavitsky} and are neglected here.) Under such circumstances the dynamical maps $\mathcal{M}_{1,2}(t)$ would be unitary, that is
\beq
\mathcal{M}_1(t)\otimes\mathcal{M}_2(t)\proj{S}= \\ U_1(t)\otimes U_2(t)\; \proj{S}\; U_1^\dag(t)\otimes U_2^\dag(t).
\eeq
Moreover, the action of the unitaries on ``their'' respective spin is equal, i.e.\ formally $U_1(t)=U_2(t)$. The singlet initial state, however, is invariant under unitary operations of the kind $U\otimes U$.
This means that without the coupling to the bath of nuclear spins, the initial electronic spin state, the singlet state, remains unchanged in the presence of an external magnetic field, that is, no dynamics takes place.
Therefore, given that the initial state of the electrons is a singlet, it is the hyperfine interaction that introduces and enables a magnetic field dependence of the electron spin dynamics in the first place. Since the nuclei act only locally on the electron spins it is not possible to implement a unitary evolution in the electronic Hilbert space alone. Rather it is \emph{decoherence} through the interaction with the nuclear spin environment as the only way of introducing a dynamics on the electron spins that start out in a singlet, which necessarily requires a loss of coherences and entanglement.
Decoherence pushes the spin state of the electrons out of the stationary singlet state and thereby ``switches on'' the dynamical effects of the external magnetic field.

\begin{figure}
	\includegraphics[width=0.625\linewidth]{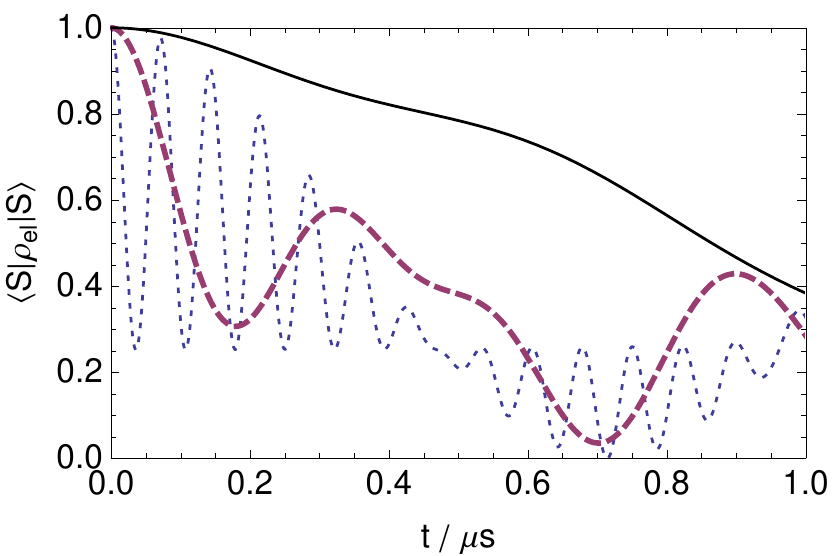}%
	\includegraphics[width=0.375\linewidth]{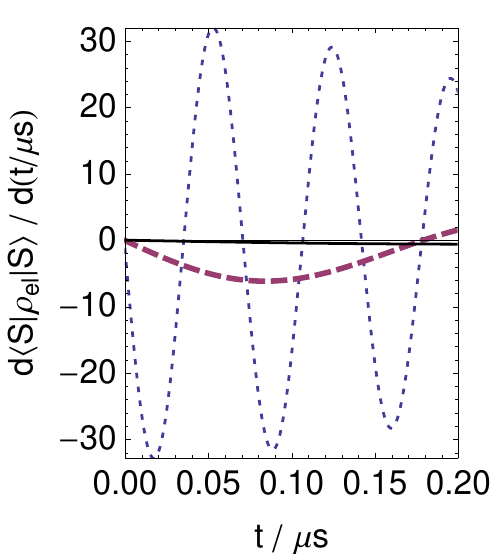}
	\caption{Dynamics of the singlet fidelity (left) and its time derivative (right) for a simple radical pair model of two electron spins and a single spin-1/2 nucleus with an external magnetic field of $B=50$\,\textmu{}T for different strengths of the hyperfine coupling: $\lambda=0.5$\,mT (dotted), $\lambda=0.1$\,mT (dashed), and $\lambda=0.02$\,mT (solid). In the limit $\lambda\to 0$, the initial state remains stationary.}
	\label{fig:smallCoupling}
\end{figure}

\begin{figure}[t]
	\begin{minipage}{0.49\linewidth}
		\includegraphics[width=\linewidth]{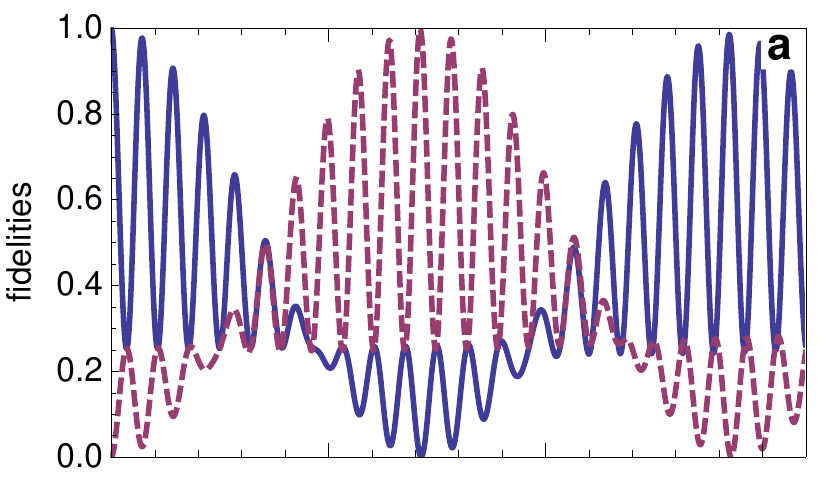}\\
		\includegraphics[width=\linewidth]{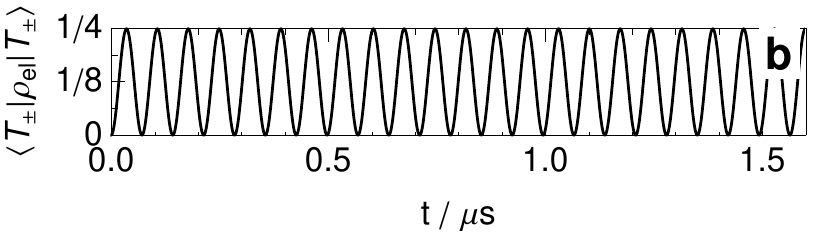}
	\end{minipage}
	\hfill
	\begin{minipage}{0.49\linewidth}
		\includegraphics[width=\linewidth]{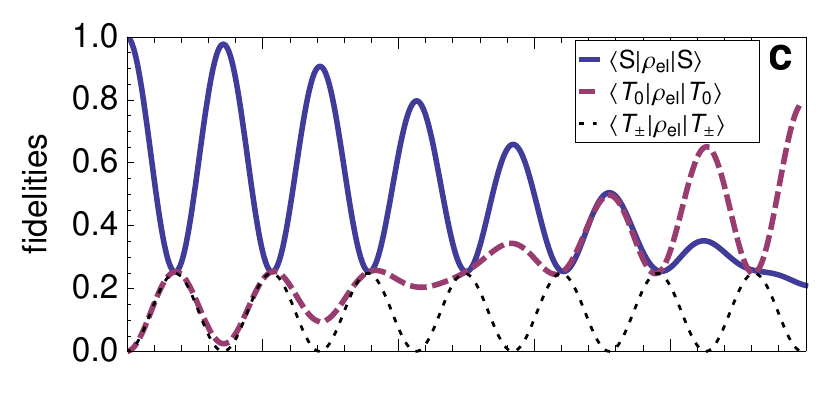}\\
		\includegraphics[width=\linewidth]{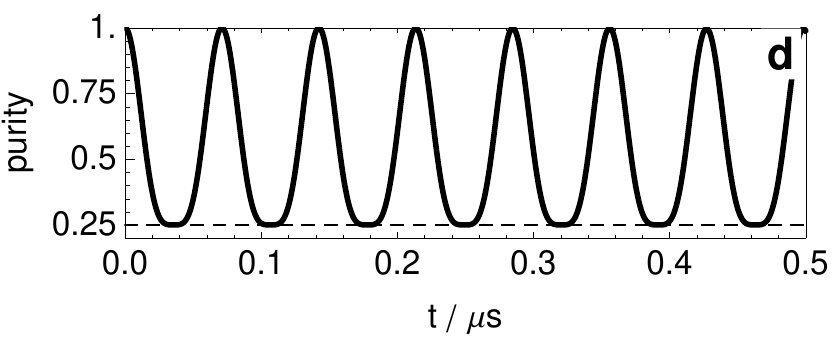}
	\end{minipage}%
	\caption{Dynamics of the electronic spin state is illustrated for the model of two electron spins and a single spin-1/2 nuclear spin coupled to one of the electrons with $\lambda=0.5$\,mT and $B=50$\,\textmu{}T. Shown are \textbf{(a)}~singlet (solid) and triplet-zero (dashed) fidelities of the electron spins, which are mixed and interconverted by hyperfine coupling and external magnetic field, and on the same time axis \textbf{(b)}~equal fidelities of the $\ket{T_\pm}$ states. \textbf{(c)}~shows the short time behavior of all the fidelities for comparison with \textbf{(d)}~the purity of the electronic spin state on the same time axis.}
	\label{fig:SimpleModel}
\end{figure}

\bigskip
Let us now illustrate these effects of decoherence in the radical pair mechanism by means of a simple model: a radical pair with only a single nuclear spin on one of the molecules with an isotropic hyperfine interaction.
This model will not capture the directional sensitivity necessary for avian magnetoreception and it will also not reproduce the spin evolution of more complex radical pairs. The simplicity of the model, however, allows us to unambiguously identify how decoherence drives the spin interconversion in principle.
The Hamiltonian of the simple model radical pair with an isotropic hyperfine interaction is given by
\beq
H = \mu_B g \left[ \vec{B}\cdot\vec{S}_1+(\vec{B}+\lambda\vec{I})\cdot\vec{S}_2 \right].
\eeq
Figure~\ref{fig:smallCoupling} shows that the dynamics of singlet fidelity of the electronic state depends on the strength of the hyperfine coupling constant $\lambda$, i.e., when viewing the dynamics as a decoherence process the coupling strength between system and environment. The smaller $\lambda$, the slower the dynamics of the spin system as indicated by a slower change in singlet fidelity. The strongest chosen hyperfine coupling $\lambda=0.5$\,mT (dotted lines) is a typical value for isotropic hyperfine couplings occurring in organic molecules, such as in Py-DMA~\cite{Rodgers2007}. The weakest chosen example $\lambda=0.02$\,mT (solid lines) is already two-fifths the magnitude of the Earth's magnetic field.
Figure~\ref{fig:SimpleModel} shows how the hyperfine interaction starts mixing the spin characters of the electron spins in addition to the interconversion driven by the external magnetic field. The singlet is partially evolved into triplet states, predominantly into $\ket{T_0}=(\ket{\up\down}+\ket{\down\up})/\sqrt{2}$ with small admixtures of the states $\ket{T_+}=\ket{\up\up}$ and $\ket{T_-}=\ket{\down\down}$. The necessary decrease in purity due to the mixing introduced by the decohering environment of nuclei, and the accompanying loss (and revival) of the singlet coherences is plotted in fig.~\ref{fig:SimpleModel}d. The minimal purity is reached for the totally mixed state $\rho_\text{el}=\one/4$ with $\Tr \rho_\text{el}^2 = 1/4$. Note that during the dynamics, also states of high purity emerge that are \emph{superpositions} of $\ket{S}$ and $\ket{T_0}$, to which the singlet has been converted.

Let us remark that these dynamics take place exclusively on many-particle coherences. The initial state of each of the spins alone is a completely mixed state, i.e.\ $\rho_\text{el 1}=\rho_\text{el 2}=\one/2$ and equally each of the nuclear spins by assumption. When observing individual spins, be it one of the electrons or one of the nuclei, they all remain in a completely mixed state, which has no coherence for a single spin.
Therefore all the spin dynamics in the radical pair involves genuine multi-particle coherences.

\begin{figure}[t]
	\includegraphics[width=\linewidth]{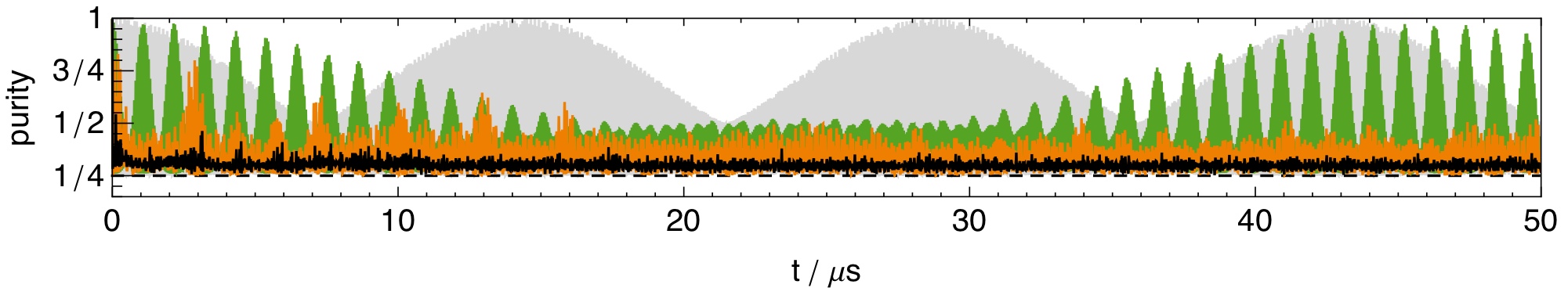}\\
	\includegraphics[width=\linewidth]{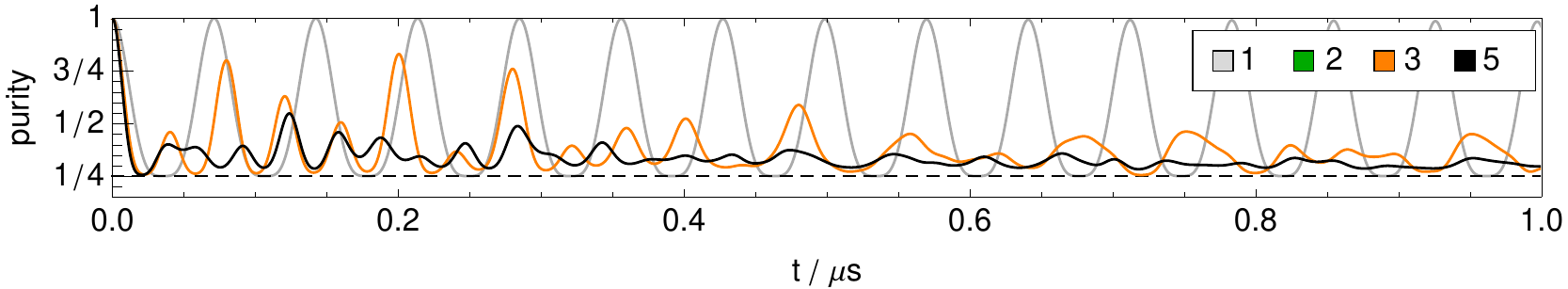}
	\caption{Dynamics of the purity of the electronic spin state for a model radical pair with different number of nuclei coupled to one of the electrons:
		\textbf{(Top)} long-time behavior for 1, 2, 3, and 5 nuclei. Revivals and thus high values of the purity are present for 1 and 2 nuclei.
		\textbf{(Bottom)} short-time behavior for 1, 3, and 5 nuclei.
		Hyperfine couplings $\lambda_{1,\dotsc,5}$ for the included nuclei are set to arbitrary values of typical magnitude with values in the order:
		0.501\,mT, 0.453\,mT, 0.376\,mT, 0.264\,mT, 0.296\,mT.
		}
	\label{fig:manySpins}
\end{figure}

The simple model with only one nuclear spin shows that the initial state re\-emer\-ges at some time given by the single hyperfine coupling and the external magnetic field as visible in fig.~\ref{fig:SimpleModel}a. A single nucleus clearly provides only a very small ``environment''.
In a different terminology one can say that the environment has a limited information capacity. The dynamics explore the relatively small Hilbert space quickly, and therefore take the system back close to its initial state quickly as well. The information that is lost from the system to the environment due to decoherence is fed back by the small environment.
However, as soon as the spectral density of the environment becomes denser, the Poincaré recurrence time, that is, the time after which the state comes back arbitrarily close to its initial state, will grow with the number of additional different nuclear spins. Each additional nucleus with a different hyperfine coupling causes the revival time to be delayed farther. This fact is illustrated in fig.~\ref{fig:manySpins}, where one spin after another with a different coupling constant is added to the environment of one electron in the model. While the electronic spin state is taken back to a pure state for one and two nuclei, this revival of the many-particle coherence is already absent on the same time scale for slightly larger environments of three and five nuclei, respectively. The close-up of the purity at short times shows that for larger environments decoherence occurs faster and that recurrences of coherences become smaller and are rarer. The principle observation of the simple model with one nucleus, namely how decoherence due to the nuclear spin environment drives the spin state interconversion, persists also for models with more nuclei. The additional effect of irreversibility is added due to the larger number nuclei.
The model of a molecule with five nuclei comes qualitatively close to the possible candidate molecules for avian magnetoreception, i.e., the flavin cofactor of cryptochrome, which has five dominant hyperfine couplings, and a model of which has been numerically investigated in~\cite{Cintolesi}.
It becomes apparent that for a relatively small number of nuclei the memory of the environment becomes big quickly, and that a transient back-flow of information to the system as measured by an increase in purity is suppressed. Revival times extend quickly to time scales that are far beyond the relevant time scales on which the chemical reactions take place. Although the nuclear spin environments are relatively small, with increasing number of spins they quickly exhibit the behavior of an environment that can practically be considered \emph{irreversible}. Even at small numbers, the nuclear spins of the radical molecules come close to the notion of an infinite \emph{bath} or ``the rest of the universe''.

While here we restrict to the analysis of purity and fidelities for a simple model, the magnetic field dependence of entanglement has been analyzed for Py-DMA in~\cite{Cai2010}. Entanglement as a hallmark of quantumness of a composite system is a stronger form of multiparticle coherence. For entanglement to exist, coherences between subsystems are required, but the mere presence of coherences is not sufficient for subsystems to be entangled. In~\cite{Cai2010} entanglement lifetime, which is defined as the timescale that decoherence needs to break entanglement, is used as a novel signature of the magnetic field dependence in the radical pair mechanism.
It also provides a measure for the time that decoherence needs to turn quantum into classical correlations.

\bigskip
From the perspective of decoherence studies one can therefore conclude that decoherence is an essential mechanism that drives the chemical compass as it is supposed to work in migrating birds. Here, we have restricted ourselves to isotropic hyperfine interactions and thus to a scenario that provides a magnetometer, which is only dependent on the \emph{strength} of the magnetic field and not its direction. For a compass, it is necessary to also consider anisotropic hyperfine interactions in order to gain a dependence of the spin dynamics on the \emph{direction} of the magnetic field. Nonetheless, our arguments also apply in the compass scenario because they do not rely on the specific form of the hyperfine interaction. It is therefore the environment of nuclear spins in the molecule(s) that constitute the radical pair in the bird compass, which alters the many-particle coherences of the electron spins and thereby introduces dynamics in the first place. It is thus the decoherence that introduces a dependence of the chemical compass on an external magnetic field and thus plays a constructive role.
An attempt to ``protect'' the electron spins from this decoherence by applying dynamical decoupling techniques known from quantum control, i.e., techniques that coherently drive the system by external means on the quantum level, will in fact result in a decreased functionality of the chemical compass or even turn it off completely~\cite{Cai2010}.

After having discussed the dynamics of the electronic spin state, we now proceed with the subsequent step in the radical pair model, which is responsible to convert the information contained in the spin state into a usable signal.

\section{Recombination mechanism}

After the radical pair has been created from a light-induced electron transfer, and the spin state of the radicals have evolved, the last step of the radical pair mechanism is the spin-dependent recombination of the electrons.
The spin-dependence of this step provides the possibility to convert microscopic properties of the spin state, e.g.\ the amount of singlets, into a macroscopic signal that can be chemically or physically detected easily.
For example, for fluorescing radical pairs in solution, a singlet spin state enables the recombination of the electron to the electronic ground state by emitting a light quantum. A triplet state however, would remain dark. Due to the initial light activation of the radical, spin states that are separated by a minuscule amount of energy, small enough to be interconverted into one another by hyperfine and Zeemann interaction (see above), are amplified to quantum states that differ in energy by much more ($\sim$1\,eV), e.g.\ fluorescence light vs.\ no light.

In this section, we are concerned with the process that converts the spin state into reaction products of ``classical character'', that is, either one of the outcomes occurred, but not a coherent superposition of them.
One may therefore be reminded of a measurement apparatus that gives one out of two possible classical measurement results after ``measuring'' the spin state at a certain time. In this context we are again confronted with decoherence. For a complete description of the recombination mechanism, the ``measurement apparatus'' needs to be taken into account within a quantum description. Thereby another environment, the apparatus, interacts with the radical spins and introduces another source of decoherence. Eventually this source of decoherence will lead to a incoherent spin mixture of either singlet or triplet, and the corresponding respective reaction product. However, since this process happens on a finite time scale, it may for particular realizations also affect the dynamics of the spin evolution \emph{while} the spins are evolving.

For radical pair molecules in solution, one can first and foremost think of the mole\-cules being close together for the initial electron transfer, followed by a time of separation, in which the spin state evolves while the molecules diffuse in solution. Upon their reencounter, the electrons either recombine if their spin state is a singlet, or they remain separated on the molecules, which again separate to never meet again. (Another reencounter, after an unsuccessful one will give corrections of higher order.)
Under these circumstances, a reasonably accurate and therefore generally adopted phenomenological recombination model is the \emph{exponential model}, also known as Haberkorn model~\cite{Rodgers2007,Haberkorn}.
The molecules reencounter with a certain probability after their random walk through the solution. The amount of reaction product originating from the singlet state of the radicals (e.g.\ fluorescence intensity) comes from sampling the singlet fidelity with an exponential reencounter probability.
The \emph{singlet yield} for such a reaction is
\beq \label{eq:exponentialModel}
\Phi_S = \int_0^\infty dt\; k e^{-kt} \bra{S}\rho_\text{el}(t)\ket{S}.
\eeq
The magnetic field effects contained in the dynamics of the electronic spin state are thereby imprinted into the amount of reaction product, i.e.\ we have $\Phi_S=\Phi_S(\vec{B})$.
In applications of the radical pair mechanism, the amount of reaction products is directly proportional to the singlet yield and thus generally depends on the magnetic field.
Regarding applications for sensors and magnetoreception, the quantities that measure the performance on the stage of the radical pair mechanism are the magnetic field sensitivity $\partial\Phi_S/\partial B$ and/or the directional profile and angular resolution of $\Phi_S=\Phi_S(\vartheta,\varphi)$ with $\vartheta$ and $\varphi$ being the polar and azimuthal angle of the magnetic field, respectively.

In more complex radical pair molecules (biradicals, proteins), there is no diffusion involved such that the recombination mechanism acts all the time. In these systems, singlet, as well as triplet components return to the ground state but at different rates (recombined triplet excited states mostly being long-lived). The whole process is therefore driven at a speed that is spin-dependent, which ultimately gives rise to different, spin-dependent concentrations of the radical molecules. In experiments it is the absorbance of this radical pair state that is measured and exhibits magnetic field dependences. Regarding avian magnetoreception, it could be this radical pair state that constitutes a signaling state of the involved protein, which causes the excitation of neural signals similar to light reception in the retina~\cite{Ritz00}. 

The simple exponential model in the form of~\eqref{eq:exponentialModel} does not contain a quantum description of the recombination process and neither takes into account the quantum nature of interaction with the environment, as well as its back-action on the spin system.
In general, the reaction kinetics of the electron back-reaction are modeled by a stochastic Liouville equation, which takes into account how the electron transfer occurs probabilistically dependent on the underlying microscopic dynamics, and how the non-reacted system evolves in the presence of such a recombination process. This formalism naturally gives a description on the level of the considered system degrees of freedom only by disregarding the environmental degrees of freedom that mediate the transfer processes.

The idea that a quantum measurement process in the Liouville equation, i.e.\ measurements of a quantum system including the effect on the state of the system as formulated in the axioms of quantum theory, influences the radical pair mechanism has been suggested~\cite{Kominis}, and started a discussion of its influences, if any, to recombination mechanisms in the radical pair model, see e.g.~\cite{Kominis,Jones,Ivanov,Shushin,Kominis2011,Jones2011}. Here, we discuss possible conceivable approaches within a common quantum formalism using master equations of Lindblad form as detailed in~\cite{Tiersch}. That is, we incorporate the recombination reaction and thereby the decoherence process by means of an equation of motion for the complete spin state of the radical pair. This master equation implements the dynamics under a coupling to an environment that causes the reaction step. Master equations of Lindblad form are special cases of Liouville equations that possess a number of convenient mathematical properties, similar to the original approaches by Bloch and Redfield to describe spin relaxation~\cite{Bloch,Redfield}. Lindblad equations can be derived by imposing the Born-Markov approximation, i.e.\ weak coupling and neglecting memory effects in the environment, as well as a coarse graining in time for the bath dynamics, i.e.\ the assumption that the bath degrees of freedom evolve much faster that those the involved radical spins~\cite{Breuer,Gardiner}. These assumptions are well fulfilled for quantum optical systems, which show a similar phenomenology, e.g.\ fluorescent decay.

Before we start with the specific recombination dynamics, we emphasize that the decoherence observed in this section will emerge with respect to an environment different from the nuclear spins used above. In contrast to many other physical systems, here, coherences will not be destroyed between particles or subsystems, but for the state in the specific basis of eigenstates of total spin, i.e.\ singlet and triplet states of the electrons. This is an unusual form of decoherence for many fields of physics. Although coherences are destroyed in the singlet/triplet basis when projecting to a singlet state, for example, at the same time coherences are established between the particles in their product basis. In fact, decoherence is always basis dependent.

\subsection{Spontaneous decay}

Inspired by the scheme in fig.~\ref{fig:RPM}, we start out with a set of excited states, the singlet and triplet radical pair states for which the electrons are on separate molecules, and a ground state that exhibits a singlet state for the electrons with both electrons being recombined on the donor molecule.
In order to keep the discussion transparent, we will restrict to reaction dynamics from the radical pair singlet only. Similar reaction processes can be incorporated for the triplet state in the same fashion as for the singlet.

In analogy to spontaneous decay in an excited atom, a first approach to the reaction mechanism is to model by means of a spontaneous decay from the singlet radical pair state to the ground state. This process requires a coupling to the environment that releases the energy difference. For fluorescing radical pairs in solution, the environment is indeed the radiation field to which the energy is released in form of a photon. One interaction Hamiltonian of the molecular system and the radiation field could be
\beq
H_{int} = g (a^\dag \otimes L_S + a\otimes L_S^\dag),
\eeq
where $g$ is the coupling strength, and $a$ and $a^\dag$ are the creation and annihilation operators of the relevant electromagnetic field mode that take and release a photon from/to the mode, respectively. These operations occur together with operators $L_S=\ket{P}\bra{S}$ and $L_S^\dag$ that act on the molecular system. $L_S$ transforms the radical pair state with electrons in a singlet $\ket{S}$ to the product state $\ket{P}$, in which both electrons are recombined on the donor molecule in the singlet ground state.

For these reaction dynamics, and assuming that the environment is initially in the vacuum ground state, the Lindblad master equation is
\beq
\frac{\partial\rho(t)}{\partial t} = -\frac{i}{\hbar} [H,\rho] + k_S \left(L_S \rho L_S^\dag - \frac{1}{2} \left\{L_S^\dag L_S,\rho\right\}\right),
\eeq
where in addition to the coherent evolution of the electron and nuclear spins only the incoherent spontaneous decay from the singlet radical state occurs with rate~$k_S$. The expression $\{A,B\}=AB+BA$ denotes the anticommutator. A similar equation is used in~\cite{Gauger}.

This equation guarantees that the singlet part of a coherent superposition in the singlet/triplet subspace of the radical pair decays exponentially with rate $k$ into the product state. Since we have traced over the environment, e.g.\ the field mode, there are no coherences between the radical pair state and the recombined ground state. In addition, due to the anticommutator term, the coherences between the singlet state and all of the triplet states of the unrecombined radical pair decay at the rate $k/2$, what recovers the original Haberkorn model~\cite{Haberkorn,Ivanov}. In effect, this appears to be an additional decoherence source for the radical pair spin state. It destroys coherences of the unrecombined radical pair in the singlet/triplet basis as it takes out singlets from the state.
Note, that in contrast to local decoherence processes, coherences are not destroyed between particles, i.e.\ between the electron spins in the product basis $\{\ket{\up\down}, \ket{\down\up}, \dots\}$, but in the basis of the correlated singlet and triplet states. This is possible because the decay involves the global operation on both electrons that extracts the singlet fraction. A triplet state would not be affected by the decay. The coherent dynamics due to $H$, however, will slowly start converting and mixing triplets to singlet states.

If we would add a similar decay for the states of the triplet manifold, e.g.\ a process that would remove the triplets at the same rate, then we recover the simple exponential model. There, the entire radical pair subspace decays uniformly at the same rate, i.e.\ without affecting the coherent spin dynamics and the rate at which the spin character changes. In this special case, the singlet yield is given by~\eqref{eq:exponentialModel}.

\subsection{Pure dephasing}

In contrast to a recombination dynamics via a spontaneous decay, there are external degrees of freedom that may not cause the recombination but only affect coherences between singlets and triplets.
For example, once the radicals and hence the electrons come close to one another, dipolar and exchange interaction can no longer be neglected. The exchange interaction, for instance, separates the electron singlet and triplet states in energy. For radical pairs in solution, the different spin states thus cause an attractive or repulsive potential, respectively. The spin state of the radicals is thereby coupled to the classical molecule motion in solution.
On one hand this constitutes a measurement of the spin character, with the molecule path being the ``pointer'' of the measurement apparatus. The molecules thus take a slightly different path depending on the spin character. On the other hand, this causes a dephasing, a decay of coherence in the singlet/triplet basis of the radical pair~\cite{Shushin}.
A possible interaction Hamiltonian between the radical pair spins and such an environment, e.g.\ the molecule position, is
\beq
H_{int} = g \left( -p\otimes P_S + p \otimes P_T \right).
\eeq
The different spin characters, captured by the projection operators $P_S=\proj{S}$ and $P_T=\proj{T_0}+\proj{T_+}+\proj{T_-}$ onto the singlet and triplet subspaces, respectively, are thereby coupled with strength $g$ to (Hermitian) operators $\pm p$ of the environment, e.g.\ momentum kicks.
Note that in contrast to the spontaneous decay, now also the triplet states are explicitly coupled to the environment.
The dynamics of the radical pair is then given by the Lindblad master equation
\beq
\frac{\partial\rho(t)}{\partial t} = -\frac{i}{\hbar} [H,\rho] + \sum_{i=S,T} k \left( P_i \rho P_i - \frac{1}{2} \left\{P_i,\rho\right\}\right).
\eeq
This form of decoherence only destroys the coherences between singlet and triplet states at rate~$k$. It thereby slows down the interconversion of singlets and triplets because here both spin characters are indeed measured by the environment and thus projected back into an incoherent, probabilistic mixture of them. For a strong interaction, i.e.\ sufficiently large $k$, one obtains a quantum Zeno effect that slows down the spin dynamics~\cite{Kominis}.
The measurement result of the spin character, however, may be practically inaccessible because the change to the environment is small and the information is quickly distributed over the environment, e.g.\ due to Brownian molecule motion, the microscopic causes of the path taken are quickly lost. In addition, the involved energy scales are much smaller as compared to the case when energy is released due to recombination.

\subsection{Quantum measurement induced recombination}

Finally, let us combine both of the previous mechanisms into a more complex recombination procedure. Now, an environment is assumed to first couple to the radical pair spin states in a way that effectively results in a measurement of the spin character. Depending on the measurement outcome, the radical pair recombines if the measured spin state was in a singlet. For the triplet nothing happens.
A possible interaction Hamiltonian that causes these combined dynamics is
\beq
H_{int} = g_S \left( \tilde{a}^\dag \otimes L_S + \tilde{a}\otimes L_S^\dag \right) + g_T p \otimes P_T,
\eeq
where the part of measuring the singlet my means of $p\otimes P_S$ and the successive decay under emission of the energy difference, $a^\dag\otimes L_S$, have been combined into a modified version of the latter for brevity of the expressions.
The corresponding Lindblad master equation therefore reads
\begin{align}
\frac{\partial\rho(t)}{\partial t} =& -\frac{i}{\hbar} [H,\rho] \nonumber \\
&+ k_S \left( L_S \rho L_S^\dag - \frac{1}{2} \left\{L_S^\dag L_S,\rho\right\}\right)
+ k_T \left( P_T \rho P_T - \frac{1}{2} \left\{P_T,\rho\right\}\right).
\end{align}
It is apparent that the master equation contains the respective parts of the processes above.
The first incoherent expression contains the spontaneous decay after a detected singlet, and
the second expression the measurement of the triplet character.
In contrast to the spontaneous decay, now also the triplet is measured and the environment obtains information about the triplet. Although, here, a measurement of the triplet does not cause a recombination, the mere fact that the environment obtains information about the presence of the triplet causes a different dynamics.
Not only is the singlet extracted at a rate $k_S$ and thus the singlet-triplet coherences decay at rate $k_S/2$, the detection of the triplet causes an additional increase of the decay rate of the coherences by $k_T/2$ to $(k_S+k_T)/2$.

The mere detection of the triplet character by the environment causes an additional decoherence source similar to the one described in pure dephasing. In addition to the decay of the singlet, the radical spins are under the influence of an effective dephasing that again slows down the spin interconversion.
When projected onto the subspace of singlet and triplet one obtains the Jones-Hore reaction operator~\cite{Jones}.
The resulting dynamics of the quantum measurement induced recombination amounts to a more general dynamics that also contains the spontaneous decay in the limit of setting $k_T=0$, a vanishing rate of the additional triplet measurement.

\bigskip
In this section, we have summarized possible sources of decoherence for the recombination step of the radical pair mechanism, in addition to the decoherence processes identified for the coherent spin dynamics in the previous section.
For the reaction, on one hand, a decay-like dynamics is required in order to release excess energy to the environment and form the reaction product. This step constitutes an amplification of the properties of the radical spin state by several orders of magnitude in energy, and therefore is an essential step to convert the properties of a quantum state to the macroscopic world.
Within this amplification step, the environment that surrounds the radical pair molecules may be seen as a measurement apparatus, which performs measurements on the radical pair spins and produces the reaction products as measurement outcomes.
On the other hand, this measurement of the spin character destroys coherences in the spin systems of the radical pair. If information about the spin character is obtained at a rate greater than at which the product is produced, e.g.\ by an additional measurement of the triplet character, an additional decoherence source on the spin state appears, which slows down the coherent spin dynamics~\cite{Kominis}. Measurement of the spin character thus may additionally influence the spin dynamics and hence the function and performance of the chemical compass.

The avian magnetoreception hypothesis in birds relies crucially on the fact that quantum states of spins are amplified to biochemical energy scales that can be physiologically and neurally processed further to establish a sense for magnetic field directions. This form of decoherence is a vital ingredient for the reaction in the sensory process.

\section{Conclusions}

We have identified that from the perspective of decoherence studies, the radical pair mechanism, one of the main hypothesis for the magnetoreception of birds, is influenced and relies on several sources of decoherence at various steps within the mechanism. That is, while spin interconversion in the radical pair model is understood in detail, and there exist various suggestions for the reaction operator describing the electron recombination, we supply the novel perspective to look at these processes from the viewpoint of decoherence by using the language of open quantum systems. Thereby, we identify several places where a decoherence processes can be identified as the driving mechanisms of the radical pair model.
First and foremost we found that decoherence of the electronic spin state due to an environment of nuclear spins enables the functioning of the radical pair mechanism and renders the process sensitive to magnetic fields.
Decoherence initiates the spin dynamics of the radicals.
Second, decoherence is required to transform and amplify quantum states of microscopic degrees of freedom, spins, to physiological scales and thus bridges the quantum to classical transition.
Thereby decoherence renders spin degrees of freedom useable for controlling and influencing chemical and physiological processes.
A direct experimental observation of the exact decoherence dynamics would be very challenging and ideally require single-electron spin readout in single molecule experiments. Concerning the decoherence of the electronic spin states due to the environment of nuclei in particular, an indirect modification of the decoherence dynamics could be achieved by modifying or even removing the nuclear spin environment in the molecules by isotope substitution.

It seems that the radical pair mechanism provides an instructive example of how the behavior of macroscopic entities, like the European robin, may indeed remain connected, in an intriguing way, to quantum processes on the molecular level.


\section*{Acknowledgments}
The research was funded in part by the Austrian Science Fund (FWF): F04011, F04012.


\end{document}